\documentclass[twocolumn,aps,superscriptaddress,prb]{revtex4-1}
\usepackage{graphicx}
\usepackage{amssymb}
\usepackage{amsmath}
\usepackage{bm}
\usepackage{color}

\usepackage{braket}
\usepackage{siunitx}
\usepackage[top=30truemm,bottom=30truemm,left=25truemm,right=25truemm]{geometry}

\def\Vec{\mathbf}

\def\lsim{\, \lower -0.3ex \hbox{$<$} \kern -0.75em \lower 0.7ex \hbox{$\sim$} \,}
\def\gsim{\, \lower -0.3ex \hbox{$>$} \kern -0.75em \lower 0.7ex \hbox{$\sim$} \,}

\begin{document}

\title{Diamagnetic levitation and thermal gradient driven motion of graphite}
\author{Manato Fujimoto}
\affiliation{Department of Physics, Osaka University,  Osaka 560-0043, Japan}
\author{Mikito Koshino}
\affiliation{Department of Physics, Osaka University,  Osaka 560-0043, Japan}
\date{\today}

\begin{abstract} 
We theoretically study the diamagnetic levitation and the thermal-driven motion of graphite.
Using the quantum-mechanically derived magnetic susceptibility,
we compute the equilibrium position of levitating graphite over a periodic arrangement of magnets,
and investigate the dependence of the levitation height on the susceptibility and  the geometry.
We find that the levitation height is maximized at a certain period of the magnets,
and the maximum height is then linearly proportional to the susceptibility of the levitating object.
We compare the ordinary AB-stacked graphite and a randomly stacked graphite,
and show that the latter exhibits a large levitation length particularly in low temperatures,
because of its diamagnetism inversely proportional to the temperature.
Finally, we demonstrate that the temperature gradient moves
the levitating object towards the high temperature side,
and estimate the generated force as a function of susceptibility.
\end{abstract}

\maketitle

%%%%

\section{Introduction}

Diamagnetism is a property of material to repel a magnetic field.
Materials with strong diamagnetism can even levitate freely over a magnet,
and it is called the diamagnetic levitation.
The best known example of this is the Meissner effect of superconductors, 
while normal-state diamagnetic materials can also levitate under an appropriate experimental setup.  
%The diamagnetic levitation has long been a topic of interest in physics.
The stable levitation of graphite and bismuth was first demonstrated in 1930's.\cite{braunbek1939freies}
It was more recently shown that even a piece of wood and plastic \cite{beaugnon1991levitation}
and also a living frog \cite{berry1997flying, simon2000diamagnetic} and cell \cite{winkleman2004magnetic}
are able to levitate with a powerful magnet, due to their tiny diamagnetism.

\begin{figure}
  \begin{center}
    \leavevmode\includegraphics[width=0.6 \hsize]{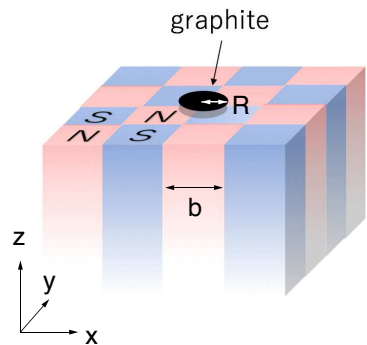}
    \caption{Diamagnetic levitation of graphite on a checkerboard array of magnets.}
    \label{fig_schem}
  \end{center}
\end{figure}

Graphite is one of the strongest diamagnetic materials among natural substances,
and its anomalous magnetic susceptibility originates from 
the orbital motion of the Dirac-like electrons.  \cite{mcclure1956diamagnetism,fukuyama1970interband,
sharma1974diamagnetism,fukuyama2007anomalous,koshino2007orbital}
The diamagnetic levitation of graphite was also extensively studied 
and various applications have been proposed. \cite{waldron1966diamagnetic, 
moser2002precise,moser2002optimization,boukallel2003levitated, cansiz2004stable, li2006lateral, liu2008variable,mizutani2012optically,kustler2012extraordinary,
kobayashi2012optical,hilber2013magnetic,su2015micromachined,kang2018design,
niu2018graphene, ewall2019optomechanical}
A typical experimental setup used for the diamagnetic levitation 
is a checkerboard arrangement of NdFeB magnet as shown in Fig.\ \ref{fig_schem},
where the alternating pattern of magnetic poles generates a magnetic field gradient
to support a diamagnetic object in a free space. \cite{moser2002optimization,kustler2012extraordinary,mizutani2012optically,kobayashi2012optical,niu2018graphene, ewall2019optomechanical}
A recent experiment performed a detailed measurement of the levitation height
of a graphite piece in this geometry. \cite{kobayashi2012optical}
The same experiment also demonstrated an optical motion control in the diamagnetic levitation, 
where the levitating graphite is moved 
towards the photo irradiated spot, motivated by the photothermal change in the magnetic susceptibility. \cite{kobayashi2012optical,ewall2019optomechanical}

In this paper, we present a detailed theoretical study of the 
diamagnetic levitation and the thermal-driven motion of graphite.
Using the orbital diamagnetic susceptibility $\chi$ calculated from the standard band model, 
we compute the equilibrium levitating position of a diamagnetic object
over the checkerboard magnet, and obtain the levitation height as a function of 
$\chi$, the checkerboard period, and the size of the object.
We find that the levitation height is maximized at a certain period of the magnets,
and the maximum height is then linearly proportional to $\chi$.
Finally we demonstrate that the temperature gradient moves
the levitating object to the high temperature side,
and estimate the generated force as a function of susceptibility.

\begin{figure}
  \begin{center}
    \leavevmode\includegraphics[width=1. \hsize]{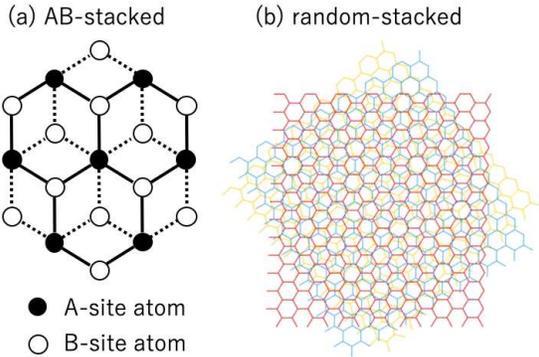}
    \caption{Atomic structure of (a) AB-stacked graphite and (b) randomly stacked graphite.}
    \label{fig:model_AB}
  \end{center}
\end{figure}

In addition to the ordinary graphite with AB (Bernal) stacking structure [Fig.\ \ref{fig:model_AB}(a)],
we also consider a randomly stacked graphite [Fig.\ \ref{fig:model_AB}(b)], in which successive graphene 
layers are stacked with random in-plane rotations. 
There the reduced interlayer coupling leads to
a strong diamagnetism inversely proportional to the temperature \cite{ominato2013orbital}, 
and therefore a large levitation length is achieved in low temperatures.
In the liquid nitrogen temperature (77K), for example, 
the maximum levitation length is found to be about 5 mm, which is 10 times as large as 
the typical levitation height of the AB-stacked graphite.

The paper is organized as follows.
In Sec.\ \ref{sec_chi}, we briefly introduce the magnetic susceptibility of AB-stacked graphite and 
randomly-stacked graphite.
We then calculate the magnetic levitation 
of general diamagnetic objects in the checkerboard magnet array in Sec.\ \ref{sec_lev}.
We consider the thermal-gradient force in magnetic levitation in Sec.\ \ref{sec_thermal}.
A brief conclusion is given in Sec.\ \ref{sec_concl}.
The susceptibility calculation for the AB-stacked graphite is presented in Appendix \ref{sec_app}.

%%%
%%%%%%%%%%%%%%%%%%%%%%%%%%%
\section{Magnetic Susceptibility}
\label{sec_chi}

We calculate the magnetic susceptibility of graphite
using the quantum mechanical liner-response formula \cite{fukuyama1971theory},
and the standard band model.
\cite{wallace1947band,mcclure1957band,slonczewski1958band,dresselhaus2002intercalation}
The detail description of the calculation is presented in Appendix \ref{sec_app}.
Figure  \ref{fig_chi}(a) plots the susceptibility $\chi$ of AB-stacked graphite
 as a function of temperature. 
Throughout the paper, we define  $\chi$ as the dimensionless susceptibility in the SI unit
 (the perfect diamagnetism is $\chi = -1$).
 In decreasing temperature, 
$\chi$ slowly increases nearly in a logarithmic manner, and finally saturate around $T \sim 50$ K.
The logarithmic increase is related to the quadratic band touching in the in-plane dispersion,
and the saturation caused by the semimetallic band structure of graphite, as argued in Appendix \ref{sec_app}.

The susceptibility of random-stacked graphite is approximately given 
by that of an infinite stack of independent monolayer graphenes.
This simplification is valid when the twist angle $\theta$ between adjacent layers is not too small ($\theta \gg 1^\circ$).
If a small twist angle happens to occur somewhere in the random stack,
these two layers are strongly coupled to form flat bands \cite{bistritzer2011moirepnas,de2012numerical}, 
and do not participate in the large diamagnetism given by the nearly-independent graphene part.
By neglecting this, the susceptibility at the charge neutral point is explicitly written as \cite{ominato2013orbital}
   \begin{equation}  
   \chi=\frac{-1}{1+k_B T/\Delta},
    \end{equation}
  where $\Delta$ is a characteristic energy scale defined by
     \begin{equation}  
%     \Delta=\mu_0 \frac{g_v g_s}{24\pi}
% \frac{e^2v^2}{d} \approx 0.03 \ \mbox{meV},
   \Delta=\frac{g_v g_s}{6}
   \left(\frac{v}{c}\right)^2
   \frac{e^2}{4\pi \epsilon_0 d} \approx 0.03 \ \mbox{meV},
    \end{equation}
 and $c$ is the light velocity.
The susceptibility is nearly proportional to $1/T$ in $k_B T \ll \Delta$ ($T \gg 0.35$ K).
As plotted in Fig.\ \ref{fig_chi}(b), the susceptibility of random-stacked graphite
is much greater than that of AB-stacked graphite particularly in the low-temperature regime.
The real system should have some disorder potential,
and then we expect that $\chi$ saturates at $k_B T \sim \Gamma$,
where $\Gamma$ is the broadening broadening near the Dirac point of graphene.

%%%%%
 \begin{figure}
  \begin{center}
    \includegraphics[width = 0.9 \hsize]{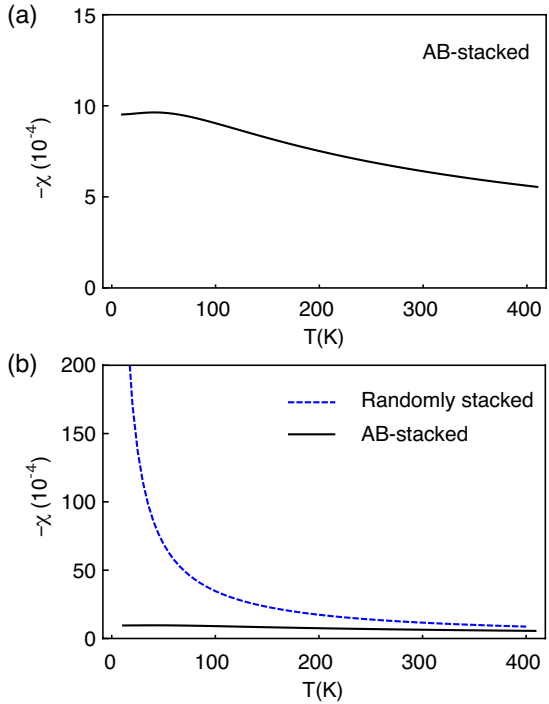}
        \caption{
        (a) Temperature dependence of the total susceptibility $\chi$ 
of AB-stacked graphite with $\mu =0$.
(b) Similar plot for the randomly stacked graphite (blue dashed),
compared to that of AB-stacked graphite (black solid).
}
    \label{fig_chi}
  \end{center}
\end{figure}

%%%%%%%%%%%
\section{Magnetic Levitation}
\label{sec_lev}

\begin{figure}
  \begin{center}
  \includegraphics[width=\hsize]{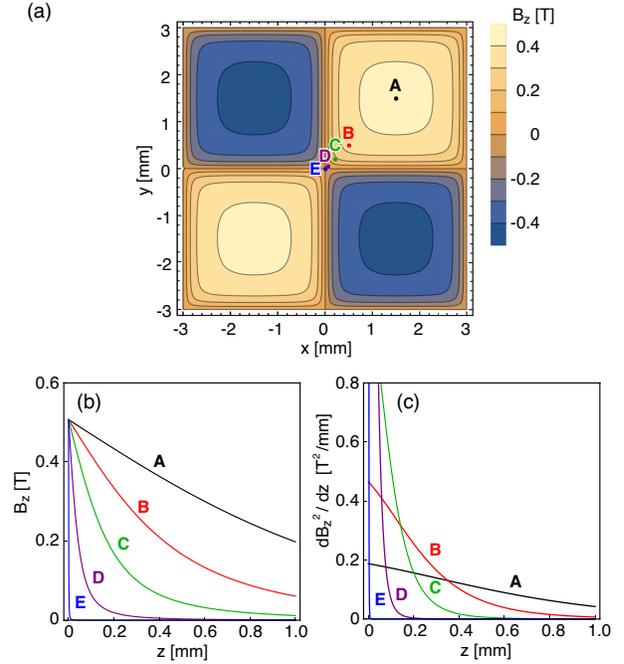}
\caption{ (a) Distribution of $B_z$ on $xy$-plane, at $0.2$ mm height 
over an infinite checkerboard magnet array with $b=3$ mm.
(b) Plot of $B_z$ as functions of $z$,
at the five points $A=(1.5,1.5), B=(0.5,0.5), C=(0.2,0.2),D=(0.05,0.05)$ and $E=(0.001,0.001)$ (in units of mm)
which are indicated in the panel (a). (c) Similar plot for $dB_z^2/dz$.
}
    \label{fig_geometry}
  \end{center}
\end{figure}

We consider magnetic levitation of graphite in the geometry illustrated in Fig.\ \ref{fig_geometry}(a).
Here the N-pole and S-pole of square-shaped magnets of size $b$ are alternately arranged
in a checkerboard pattern. We place a round-shaped graphite disk of radius $R$ and the thickness $w$ right above a grid point where four magnet blocks meet.
We assume the surface of the magnet top and the graphite disk is perpendicular to the 
gravitational direction, $z$.
The graphite is attracted to the grid point where the magnetic field is the weakest.

When the magnetic field distribution $\Vec{B}=\Vec{B}(x,y,z)$ is given,
the total energy $U$ of the graphite disk is given by
   \begin{equation}
   \label{eq:Energy}
U= M g z + w \int_S dx dy \frac{-1}{2\mu_0}\chi B_z(x,y,z)^2,
 \end{equation}
Here $g$ is the gravitational acceleration, $M$ is the mass of the disk, 
$S$ is the area of the disk, $\chi$ is the magnetic susceptibility of graphite, 
% $\mu_0$ is the permeability in vacuum, 
and $z$ is the vertical position of the disk. We assumed the thickness of graphite is thin enough.
The equilibrium position $z=z_{\rm lev}$ (i.e., the levitation length) is obtained by 
solving $\partial U /\partial z = 0$, or
 \begin{equation}
%\frac{w}{2\mu_0}(-\chi) \int_S dx dy \frac{dB_z^2}{dz} = M g.
\left\langle \frac{dB_z^2}{dz} \right\rangle_S  = 2\mu_0 \frac{\rho g}{\chi}.
\label{eq_equilibrium_position}
\end{equation}
Here $\rho \approx 2.2$ g/cm$^3$ is the mass density of graphite, and $\langle \cdots \rangle_S$ is the average over the graphite area.
For the magnetic levitation, therefore, the squared magnetic field gradient $dB_z^2/dz$ matters
more than the absolute field amplitude itself.

Now we consider an infinite checkerboard arrangement
of square blocks of NdFeB magnet.
The $z$-component of the magnetic field generated by a single block can be
calculated by the formula, \cite{camacho2013alternative}
\begin{align}
& B^{(1)}_z(x,y,z)=
-\frac{B_0}{2\pi}[F_1(-x,y,z)+F_1(-x,y,-z)\nonumber\\
& +F_1(-x,-y,z)+F_1(-x,-y,-z)+F_1(x,y,z)\nonumber\\
& +F_1(x,y,-z)+F_1(x,-y,z)+F_1(x,-y,-z)],
 \end{align}
with
 \begin{align}
 &F_1(x,y,z)= \nonumber\\
& 
\arctan
\frac
{\left(x+\frac{b_x}{2}\right)\left(y+\frac{b_y}{2}\right)}
{\left(z+\frac{b_z}{2}\right)
\sqrt{\left(x+\frac{b_x}{2}\right)^2+\left(y+\frac{b_y}{2}\right)^2+\left(z+\frac{b_z}{2}\right)^2}
},
 \end{align}
where $B_0 $ is the amplitude of the surface magnetic field,
and $b_x$, $b_y$ and $b_z$, are the side lengths,
and the $N$ and $S$ poles of the magnet correspond to the faces of $z=b_z/2$  and $-b_z/2$, respectively.
The total magnetic field $B_z(x,y,z)$ is obtained as an infinite sum of $B^{(1)}_z$
over all the blocks composing the checkerboard array.
We take $B_0=500$ mT as a typical value for NdFeB magnet, and assume
the square and long shape, i.e.,  $b_x=b_y \equiv b$, and $b_z \to \infty$.

%%%%%%%%%%%%%%%%%%%%%%%%%%%

%%%%

Figure \ref{fig_geometry}(a) shows the actual distribution of $B_z$
at $ 0.2$ mm height from the surface for $b=3$ mm magnet array.
Figures \ref{fig_geometry}(b) and (c) are the plots of $B_z$ and $dB_z^2/dz$ as functions of $z$, respectively,
at the five points $A=(1.5,1.5), B=(0.5,0.5), C=(0.2,0.2),D=(0.05,0.05)$ and $E=(0.001,0.001)$ in units of mm.
Here the coordinate origin is taken to a grid point on the upper surface of the arranged magnets.
At any $xy$-points, the magnetic field $B_z$ exponentially decays in $z$,
and its decay length is shorter when closer to the origin, and so does $dB_z^2/dz$. 

The averaged squared magnetic field $\langle B_z^2 \rangle_S$ can also be well approximated
by an exponential function in $z$ as,
\begin{align}
\langle B_z^2 \rangle_S  \approx \alpha B_0^2 e^{-z/\lambda},
\label{eq_bzsq}
\end{align}
where $\alpha$ is the dimensionless constant of the order of 1, 
and $\lambda$ is the length scale determined by the geometry.
 Then Eq.\ (\ref{eq_equilibrium_position}) is explicitly solved as
\begin{align}
z_{\rm lev} \approx \lambda \log \frac{\alpha\lambda_0}{\lambda}
\label{eq_z_lev_approx}
\end{align}
with the characteristic length,
\begin{align}
\lambda_0 = \frac{|\chi|B_0^2}{2\mu_0 \rho g}.
\label{eq_lam0}
\end{align}
The negative solution of Eq.\ (\ref{eq_z_lev_approx})
indicates that the graphite does not levitate.

\begin{figure}
  \begin{center}
    \includegraphics[width=0.85\hsize]{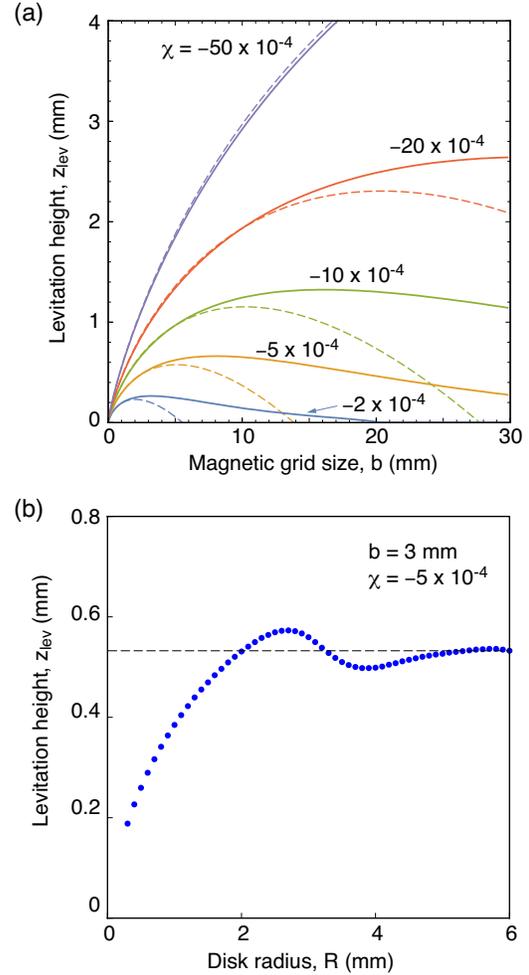}
    \caption{
(a)  Levitation height $z_{\rm lev}$ as a function of the magnetic grid size $b$,
calculated for different $\chi$'s with the limit of $R \gg b$.
Solid curves are the numerical solution of Eq.\ (\ref{eq_equilibrium_position}),
and the dashed curves are the approximate expression Eq.\ (\ref{eq_z_lev_approx}).
(b) $R$-dependence of the levitation height in $\chi = -5 \times 10^{-4}$
and the magnetic grid with $b=3$ mm.
The horizontal dashed line indicates the asymptotic value in $R\to \infty$.
}
        \label{fig_chi_vs_lev}
  \end{center}
\end{figure}

If the graphite radius $R$ is much greater than the magnet grid size $b$, in particular,
$\langle B_z^2 \rangle_S$ is replaced by the average value over whole $xy$-plane,
and then $\lambda$ depends solely on $b$ (not on $R$).
In this limit, we have $\alpha \sim 0.7$ and $\lambda  \sim \eta b$ with $\eta \sim 0.11$.
Figure \ref{fig_chi_vs_lev}(a) plots the levitation height $z_{\rm lev}$
as a function of the grid size $b$, calculated for different $\chi$'s in this limit,
The solid curves are the numerical solution of Eq.\ (\ref{eq_equilibrium_position}),
and the dashed curves are the approximate expression Eq.\ (\ref{eq_z_lev_approx})
with $\alpha=0.7$ and $\eta=0.11$.
The curves with different $\chi$'s are just scaled through the length parameter $\lambda_0$.
Here we take $-\chi = 2,5,10,20,50 (\times 10^{-4})$ ($\rho$ is fixed),
which give $\lambda_0$ is $0.9, 2.2, 4.5, 9.0, 22.4$mm, respectively.
As is obvious from its analytic form,
the approximate curve peaks at $b = \alpha \lambda_0 /(\eta e) \approx 2.3 \lambda_0$
($e$ is the base of the natural logarithm),
where the levitation height takes the maximum value,
\begin{align}
z^{\rm (max)}_{\rm lev} = \frac{\alpha \lambda_0}{e} \approx 0.26 \lambda_0.
\end{align}

We see that the approximation fails for $b$ greater than the peak position.
This is because Eq.\ (\ref{eq_bzsq}) is not accurate $z < \lambda$,
where the actual $\langle B_z^2 \rangle_S$ becomes higher than the approximation.
The gradient $d\langle B_z^2 \rangle_S/dz$ at $z=0$ is given by $-B_0^2/(\eta' b)$ with $\eta' \sim 0.05$,
and it gives the vanishing point of $z_{\rm lev}$ at $b= \lambda_0 / \eta' \sim 20\lambda_0$.
This is much further than the end of the approximate curve,  $b = \alpha \lambda_0 /\eta \approx 6.2 \lambda_0$.
The approximate formula is still useful for qualitative estimation of the maximum levitation length $z^{\rm (max)}$.

The important fact is that all the length scales of the system, such as the levitation height
and the grid period, are scaled by a single parameter $\lambda_0$ [Eq.\ (\ref{eq_lam0})],
which is proportional to $\chi$.
If $\chi$ is doubled, therefore, we have the same physics with all the length scale doubled.
For the typical susceptibility of AB-stacked graphite at the room temperature, $\chi = -5 \times 10^{-4}$, 
the characteristic length becomes $\lambda_0 = 2.23$ mm,
and we have $z^{\rm (max)} = 0.65$ mm at $b=7.5$ mm.
For the randomly-stacked graphite at 77K, on the other hand,
the susceptibility is about $\chi = -45 \times 10^{-4}$,
giving $z^{\rm (max)} = 5.9$ mm at $b=68$ mm.

 \begin{figure}
  \begin{center}
    \includegraphics[width=0.9 \hsize]{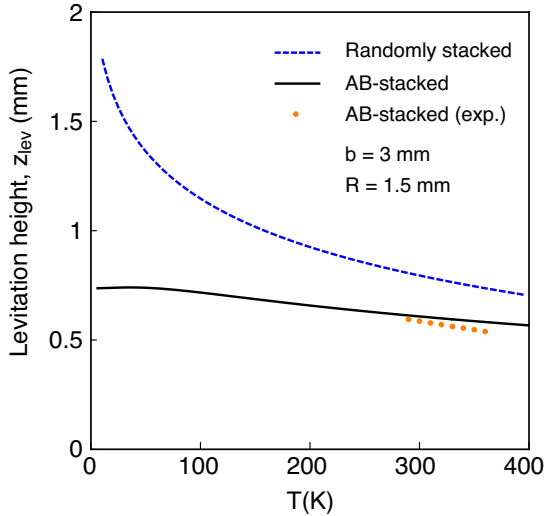}
    \caption{Temperature dependence of the levitation height of the AB-stacked graphite (blue dashed) and 
randomly-stacked graphite (black solid)
of radius $R=1.5$ mm, in the magnetic grid with $b=3$ mm.
The orange dots represent the levitation length of AB-stack graphite
measured in the experiment. \cite{kobayashi2012optical}
}
    \label{fig_lev_graphite}
  \end{center}
\end{figure}

%Figure \ref{fig_chi_vs_lev}(a) plots the levitation height $z_{\rm lev}$
%as a function of the susceptibility $\chi$, calculated for different radii $R$.
%Each single curve behaves roughly as $\propto \log (-\chi)$,
%and this can be understood by the exponential behavior of $dB_z^2/dz$ in $z$, 
%and Eq.\ (\ref{eq_equilibrium_position}).
%The curve reaches zero when $|\chi|$ is lower than a certain small value,
%and this is the minimum susceptibility for levitation.
The levitation also depends on the size of the disk.
Figure \ref{fig_chi_vs_lev}(b) shows the levitation height as a function of the disk radius $R$,
at $\chi = -5 \times 10^{-4}$ and $b=3$ mm. In increasing $R$, the levitation length first monotonically increases, 
and then eventually approaches to the asymptotic value (dashed line) argued above, after some oscillation.
The monotonic increasing region corresponds to the disk size smaller than the magnet grid size $b$.
There a smaller disk has a smaller levitation, because as seen in Fig.\  \ref{fig_geometry}(c), 
$dB_z^2/dz$ near the origin quickly decays in $z$, so
a tiny disk can levitate only in a small distance to catch the finite $dB_z^2/dz$.
%Note that the curve in Fig.\ \ref{fig_chi_vs_lev}(b) is also scaled by the length parameter $\lambda_0$.

In Fig.\ \ref{fig_lev_graphite},  we show the temperature dependence of the
levitation length $z_{\rm lev}$ of the AB-stacked graphite and randomly stacked graphite,
with a disk radius $R=1.5$ mm.
In a fixed geometry,  $z_{\rm lev}$ is proportional to $\log |\chi|$
according to Eq.\ (\ref{eq_z_lev_approx}).
We see that the  $z_{\rm lev}$ of AB-stacked graphite shows a similar temperature dependence to 
the susceptibility itself [Fig.\ \ref{fig_chi}(a)].
The randomly stacked graphite
exhibits a $\log T$ behavior because  $z_{\rm lev} \propto \log |\chi|$ and $\chi \propto 1/T$.
The orange dots in Fig.\ \ref{fig_lev_graphite} indicate the levitation length of AB-stack graphite
measured in the experiment. \cite{kobayashi2012optical}
We can see a good quantitative agreement between the simulation and experiment
without any parameter fitting.
The simulation underestimates the slope of the temperature dependence,
suggesting that the susceptibility in the real system decreases more rapidly in temperature than 
in our model calculation.
A possible reason for this would be the effect of phonon scattering, 
which increases the energy broadening $\Gamma$ in higher temperature
and reduces the susceptibility.

\begin{figure}
  \begin{center}
    \includegraphics[width=0.85 \hsize]{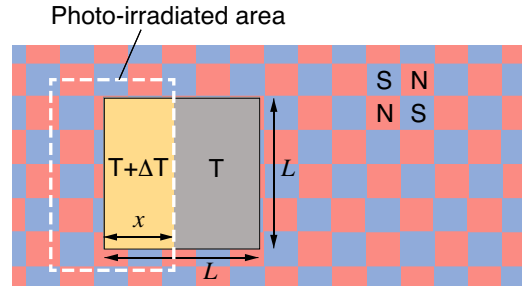}
    \caption{
    Setup of the thermal-driven motion of levitating graphite.
    A square-shaped graphite piece levitates 
    on the magnet checkerboard and it is partially heated by photo irradiation.
        }
    \label{fig_thermal_motion}
  \end{center}
\end{figure}

%%%%%%%%%%%%%
\section{Thermal gradient driven motion}
\label{sec_thermal}

A graphite piece much larger than the magnet size $b$
freely moves along the horizontal direction while floating over the magnets,
because it covers a number of magnetic periods and 
the total energy hardly depends on the $xy$ position.
Now we consider a situation illustrated in Fig.\ \ref{fig_thermal_motion},
where a part of the levitating graphite piece is heated by photo irradiation.
We assume that the irradiated area is fixed to the rest frame of the magnets,
and consider the movement of graphite against it.
In the experiment, it was shown that the graphite is attracted to the photo irradiated region.
\cite{kobayashi2012optical,ewall2019optomechanical}
This can be understood that the graphite minimizes the total energy by moving to the high temperature area,
where the diamagnetism is smaller so that the energy cost is lower under the same magnetic field.

We can estimate the magnitude of the thermal gradient force as following.
We consider a $L\times L$ square-shaped graphite piece with thickness $w$,
and assume that the graphite in the irradiated area 
is instantly heated up to temperature $T+\Delta T$,
while otherwise the temperature remains $T$, as in Fig.\ \ref{fig_thermal_motion}.
The length of the high temperature region is denoted by a variable $x$.
i.e., when the graphite is moved to the left, then $x$ increases.
We neglect the heat transport on the graphite for simplicity.
The total energy of graphite contributed by magnetic field is written as
\begin{align}
& U  = -\frac{1}{2\mu_0} \chi(T+\Delta T) \langle B_z^2 \rangle w L x
\nonumber\\
& \qquad\qquad
 -\frac{1}{2\mu_0} \chi(T) \langle B_z^2 \rangle w L (L-x),
\end{align}
where $\langle B_z^2 \rangle$ is the square magnetic field at the levitation height
averaged over $xy$-plane.
The force $f$ can be calculated as the derivative of the free energy
We can show that the free energy is dominated by the magnetic part, and then the force is obtained as
\begin{align}
f \approx -\frac{\partial U}{\partial x}  
\approx \frac{1}{2\mu_0} \frac{\partial \chi}{\partial T} \Delta T \langle B_z^2 \rangle w L.
\end{align}
For example, if we take a AB-stack graphite piece of $L=10$ 
mm and $w = 0.025$ mm, and apply a temperature difference of $\Delta T = 10$ K under $T = 300$ K, 
then we have $f = 2\times 10^{-3}$ mg-force, which gives the acceleration of 4 mm/s$^2$. 

On the other hand, we have much greater force in the random stack graphite in low temperature,
because $\partial \chi/\partial T \propto 1/T^2$.
For a piece of random-stack graphite of the same shape with $T = 77$ K
and $\Delta T = 10$ K,  the acceleration becomes 63 mm/s$^2$.

%%%%%%%%%%%%%
\section{Conclusion}
\label{sec_concl}
We have studied the diamagnetic levitation and the thermal-driven motion of graphite
on a checkerboard magnet array.
We showed that the physics is governed by the length scale $\lambda_0$ [Eq.\ (\ref{eq_lam0})],
which depends on the susceptibility $\chi$ and the mass density of the levitating object
as well as the field amplitude of the magnet.
The maximum levitation length and the required grid size 
are both proportional to $\lambda_0$, and therefore proportional to $\chi$.
We showed a randomly stacked graphite exhibits much greater levitation length than the 
AB-stacked graphite,
and it is even enhanced in low temperatures because of $\chi$ inversely proportional to the temperature.
We investigated the motion of the levitating object driven by the temperature gradient,
and estimate the generated force as a function of susceptibility.

\section*{Acknowledgments}
MK acknowledges the financial support of JSPS KAKENHI Grant Number JP17K05496.

%%%%%%%%%%%%%%%%%%%%%%%%%%%
%%%%%%%%%%%%%%%%%%%%%%%%%%%
%%%%%%%%%%%%%%%%%%%%%%%%%%%
\appendix

\section{Band Model and magnetic susceptibility of AB-stacked graphite}
\label{sec_app}

In this Appendix, we present the detailed description of the band model
and the susceptibility calculation for the AB-stacked graphite.
We consider AB(Bernal)-stacked graphite as shown in Fig.\ \ref{fig:model_AB}(a).
A unit cell is composed of four atoms, labelled $A_1$, $B_1$ on the layer 1
and $A_2$, $B_2$ on the layer 2, where $B_1$ and  $A_2$ are vertically located,
while  $A_1$ and $B_2$ are directly above or below the hexagon center of the other layer.
The lattice constant within a single layer is given by $a=0.246$ nm and 
the distance between adjacent graphene layers is $d=0.334$ nm.
The lattice constant in the perpendicular direction is $2d$.
The low-energy electronic states can be described by a ${\bf k}\cdot {\bf p}$ Hamiltonian
around the valley center $K_\pm$.
\cite{wallace1947band,mcclure1957band,slonczewski1958band,dresselhaus2002intercalation}, 
Here, we include band parameters $\gamma_i \, (i=0,1,\cdots,5)$
and $\Delta'$, where $\gamma_i$ represents the tight-binding hopping energy
between carbon atoms as depicted in Fig.\ref{fig:model_AB}, 
and $\Delta'$ is related on-site energy difference between dimer sites ($B_1,A_2$) and non-dimer sites ($A_1,B_2$) .
The parameters adopted in this work are summarized  in Table \ref{table:parameter}. \cite{charlier1991first} 

\begin{table}
 \begin{center}
 \begin{ruledtabular}
  \begin{tabular}{l l l l l l l}
    $\gamma_0$ & $\gamma_1$ & $\gamma_2$ &  $\gamma_3$ &  $\gamma_4$  & $\gamma_5$ &  $\Delta$ \\ 
    \colrule
       3.16 & 0.39 & -0.019 & 0.315 & 0.044 & 0.038 & 0.049  \\ 
  \end{tabular}
  \end{ruledtabular}
  \end{center}
\caption{Examples of the band parameters (in unit of eV) for the graphite. \cite{charlier1991first} 
%\cite{J.-C. Charlier and X. Gonze}
}
 \label{table:parameter}
\end{table}

Let $\ket{A_j}$ and $\ket{B_j}$ ($j=1,2$) be the Bloch functions at the corresponding sublattices.
If the basis is taken as $\ket{A_1},\ket{B_1},\ket{A_2},\ket{B_2}$,  the effective Hamiltonian 
is written as \cite{guinea2006electronic,partoens2006graphene,guinea2007electronic,koshino2007orbital,koshino2008magneto}
\begin{eqnarray} 
{\cal H}(\Vec{k})=
    \begin{pmatrix}
  \alpha \gamma_2 &   v p_- & -\lambda v_4 p_- & \lambda v_3 p_+  \\
 v p_+   & \alpha \gamma_5 + \Delta & \lambda \gamma_1 & -\lambda v_4 p_- \\
 -\lambda v_4 p_+ & \lambda \gamma_1 & \alpha \gamma_5 +\Delta&  v p_-   \\
 \lambda v_3 p_-  & -\lambda v_4 p_+ &   v p_+  &  \alpha \gamma_2
    \end{pmatrix}
      \label{eqn:matrix_c} 
\end{eqnarray}
where $\Vec{k}=(k_x,k_y,k_z)$, $p_\pm = \hbar(\xi k_x \pm i k_y)$, $k_x$ and $k_y$ are 
the in-plane wavenumber measured from the valley center $K_\xi$, and $\xi = \pm$ is the valley index.
We defined  $\lambda(k_z) =2 \cos{k_z d}$ and $\alpha(k_z)  =\cos{2 k_z d}$
with the out-of-plane wavenumber $k_z$.
The parameter $v = (\sqrt{3}/2) \gamma_0 a /\hbar$ is the band velocity of monolayer graphene,
and $v_3$ and $v_4$ are given by $v_i = (\sqrt{3}/2) \gamma_i a /\hbar \, (i=3,4)$.
Here $\gamma_3$ is responsible for the trigonal warping of the energy bands,
and $\gamma_4$ is for the electron-hole asymmetry.

\begin{figure}
  \begin{center}
    \includegraphics[width=0.9 \hsize]{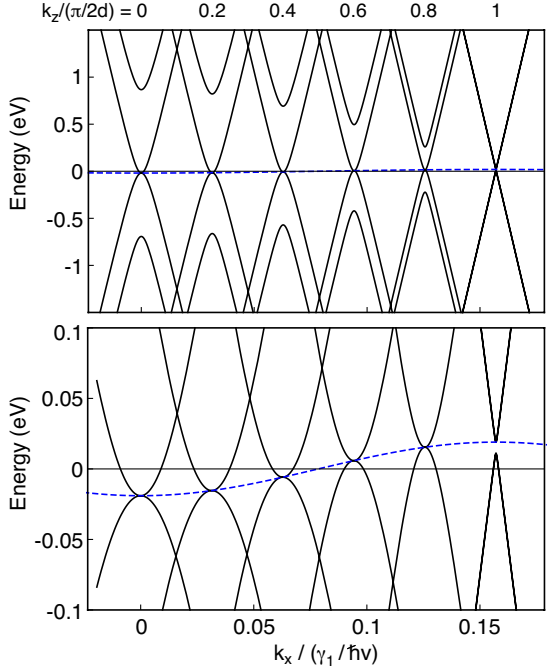}
\caption{The band structure of AB-stacked bilayer graphene as a function of $k_x$ (with $k_y=0$).
The subbands labeled by different $k_z$'s are separately plotted with horizontal shifts.
The lower panel is the magnified plot of the same bands near zero energy.
Blue dashed curves indicate the dispersion of the band touching point as a function of $k_z$.
}
    \label{fig_band}
  \end{center}
\end{figure}

Figure \ref{fig_band} shows the energy bands
as a function of $k_x$ with $k_y$ fixed to 0.
Here the subbands labeled by different $k_z$'s are separately plotted with horizontal shifts.
The lower panel is the magnified plot near zero energy.
The band structure of each fixed $k_z$ is similar to that of bilayer graphene. \cite{mccann2013electronic}
where a pair of electron and hole bands are touching near the zero energy with quadratic dispersion.
At the zone boundary, $k_z =\pi/(2d)$, 
the energy band becomes a linear Dirac cone like monolayer graphene's.
We see that the electron-hole band touching point slightly disperses in $k_z$ as $\alpha(k_z) \gamma_2$,
and this is the origin of the semimetallic nature of graphite.
 
 % Random stacked
%In a randomly stacked graphite, on the other hand,
%the momentum mismatch between the rotated layers significantly reduces
%the interlayer coupling. In that situation, we can approximately  
%treat the system as a set of independent single layer graphenes with equal spacing $d$.
%The Hamiltonian is then just given by that of monolayer graphene as \cite{ando2005theory,neto2009electronic},
%\begin{eqnarray} 
%{\cal H}(\Vec{k})=
% \begin{pmatrix}
%  0 &   v p_-  \\
% v p_+   & 0
 %   \end{pmatrix},
%\end{eqnarray}
%which is independent of $k_z$.

%%%%%%%%%%%%%%%%%%%%%%%%%%%
%%%%%%%%%%%%%%%%%%%%%%%%%%%
 \begin{figure}
  \begin{center}
     \includegraphics[width = 1. \hsize]{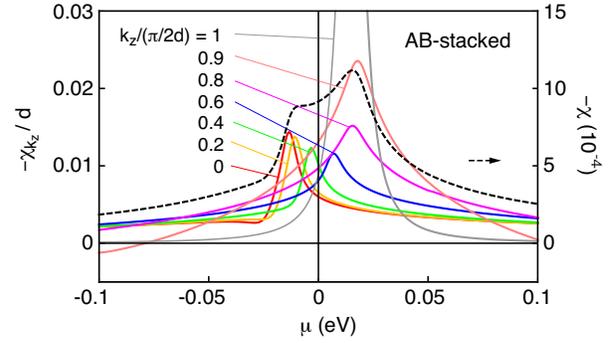}
        \caption{
Magnetic susceptibility of AB-stacked graphite  
as a function of the chemical potential $\mu$, separately plotted for different $k_z$'s. 
The dashed curve is the total susceptibility.
}
    \label{fig_chi_app}
  \end{center}
\end{figure}
%%%%%%%%%%%%%%%%%%%%%%%%%%%

For the magnetic susceptibility, we use the general expression based on the linear response theory,\cite{fukuyama1971theory}
\begin{align}
\label{eq_chi}
\chi(\mu, T)=  \int_{-\infty}^{\infty} d\epsilon f(\epsilon) \mbox{Im} F(\epsilon+i0), 
\end{align}
with
\begin{align}
F(z)=-\mu_0 \frac{g_v g_s}{2 \pi L^3} \frac{e^2}{\hbar^2} \sum_{\bf k} \mbox{tr}(G \mathcal{H}_x G \mathcal{H}_y G \mathcal{H}_x G \mathcal{H}_y).
   \end{align}
Here  $\mu$ is  the chemical potential of electrons, $T$ is
the temperature, $g_v=2$ and $g_s=2$ is the valley and spin degeneracy, respectively, 
$L$ is the system size, and $\mu_0$ is the vacuum permeability.
We also defined $ \mathcal{H}_x =\partial \mathcal{H}/\partial k_x$, 
   $ \mathcal{H}_y =\partial \mathcal{H}/\partial k_y$, $G(z)=(z- \mathcal{H})^{-1}$, 
 and $f(\epsilon)=[1+e^{(\epsilon-\mu)/k_B T} ]^{-1}$. 
 The $\chi$ of this definition is the dimensionless susceptibility in the SI unit.
 The energy density of the magnetic field is given by $-\chi B^2/(2\mu_0)$.
By integration by parts in Eq.(\ref{eq_chi}), we have
\begin{equation}
  \label{eq_chi2}
\chi(\mu, T)= \int_{-\infty}^{\infty} d\epsilon 
\left(-\frac{\partial f(\epsilon)}{\partial \epsilon} \right)  \chi(\mu, T=0),
\end{equation}
which relates the susceptibility at finite temperature with that at zero temperature.
We include the energy broadening effect induced by the disorder potential
by replacing $i0$ in Eq. (\ref{eq_chi}) with a small self-energy $i \Gamma$ in the Green's function.
We assume the constant scattering rate $\Gamma = 5$ meV in the following calculations.

Figure \ref{fig_chi_app} plots the magnetic susceptibility of AB-stack graphite
at fixed $k_z$'s (denoted as $\chi_{k_z}$)
as a function of the chemical potential with the temperature $T = 50$K,
where the dashed curve is the total susceptibility $\chi = \int^{\pi/(2d)}_{-\pi/(2d)}  \chi_{k_z} dk_z$.
%Here the panel (a) is for the minimal model with only $\gamma_0$ and $\gamma_1$,
%and (b) is for the full parameter model including all the band parameters.
Approximately, $\chi_{k_z}$ is equivalent to that of bilayer graphene,
which is a logarithmic peak centered at the band touching point
and truncated at energies of $\pm \lambda(k_z) \gamma_1$. \cite{koshino2007orbital,koshino2007diamagnetism}
In increasing $k_z$, the peak becomes higher and it finally becomes 
a broadened delta function at the zone edge $k_z=\pi/(2d)$,
which is an analog of the susceptibility of monolayer graphene. \cite{mcclure1956diamagnetism}
% All the additional feature is caused by the band parameters other than $\gamma_0$ and $\gamma_1$.
The center of the peak moves as a function of $k_z$,
in accordance with the shift of the band toughing point caused by $\gamma_2$ [Fig.\ \ref{fig_band}].
We notice that the curves near $k_z=0$ has an additional sharp peak on top of the logarithmic background, 
which originates from the trigonal warping caused by $\gamma_3$. \cite{koshino2007orbital}
The total susceptibility exhibits a broadened peak structure bound by $k_z=0$ peak
and $k_z=\pi/(2d)$ peak, and its total width is of the order of $2\gamma_2 \sim 0.04$eV.

Figure  \ref{fig_chi}(a) shows the temperature dependence of the total susceptibility $\chi$ 
of AB-stacked graphite at $\mu =0$ (charge neutral).
The temperature effect on the susceptibility can be understood by using Eq.\ (\ref{eq_chi2}),
where $\chi(\mu)$ in a finite temperature is obtained by averaging $\chi(\mu)$ of zero temperature   
over an energy range of a few $k_B T$.
In Fig.\  \ref{fig_chi}(a), $\chi$ logarithmically decreases as temperature increases,
and this is understood as thermal broadening of the logarithmic peak of $\chi(\mu)$.
When $k_BT$ is much smaller than the peak width of $\chi(\mu)$, 
the susceptibility does not depend much on the temperature, and this explains the 
nearly flat region in $T < 50$ K in Fig. \ref{fig_chi}(a).

\bibliography{graphite_lev}
\end{document}